# Invariant Vector and Majorana Star Representations of Qutrit States

Vinod Mishra

U.S. Army Research Labs, Aberdeen Proving Ground, MD 21005

*Abstract* — **The Qutrit state density matrix is of order 3 and depends on 8 parameters in general case. Visualization of this 8-dimensional state space is practically impossible using 8-dimensional vectors commonly used. Recently a 3-dimensional vector representation of the Qutrit state space (also called Invariant Vector Representation) has been proposed [1]. In this work we present the time-evolution of the IVR vectors for the Qutrit Cascade or $\Xi$—model to emphasize the advantages of the IVR and also relate IVR vectors to well-known stars of the Majorana State Representation (MSR)**

## I. Introduction

The Qutrit (three internal states) comes next in complexity after Qubit (two internal states) as a resource for quantum information processes like quantum sensing [1]. The Qubit density matrix is of order 2 and depends on 3 parameters for the most general mixed states but only on 2 parameters for pure states. It can be easily visualized using Bloch sphere representation in which the the pure states lie on the surface of the Bloch sphere and mixed states in its interior. On the other hand the Qutrit density matrix is of order 3 and it depends on 8 parameters in the most general case. Visualization of the 8-dimensional Qutrit state space is practically impossible using 8- dimensional vectors commonly used in the Gell-Mann representation.

Quantum Sensing uses quantum properties of particles and states to improve the sensitivity and accuracy of measurements beyond classical limits. Currently qubits are used as the central resource for most of this task and other ones in quantum computing and communication. Advantages of higher dimensional objects like qutrit [3,7,10] are still under investigation. One of the obstacles in this direction is the difficulties in visualization of qutrit states.

Recently another 3-dimensional representation of the Qutrit state space based on density matrix invariants (to be called IVR or Invariant Vector Representation) has been proposed [2]. These vectors also reside on the surface of a sphere and help one visualize the dynamics of a Qutrit state. In this work we apply IVR to the cascade configuration of a Qutrit (also known as the $\Xi$—model) and show its utility in understanding the dynamics of the model. We also relate it to the Majorana Star Representation (MSR) [7, 10, 11], which is used in discussing the hadronic higher spin-states.

## II. Invariant Vector Representation (IVR) of a Qutrit: Basic Results

The density matrix $\rho$ based on the spin-1 representation of a Qutrit is given as [3,5]

$$\rho = \begin{bmatrix} \omega_1 & \frac{1}{2}(q_3 + ia_3) & \frac{1}{2}(q_2 - ia_2) \\ \frac{1}{2}(q_3 - ia_3) & \omega_2 & -\frac{1}{2}(q_1 + ia_1) \\ \frac{1}{2}(q_2 + ia_2) & -\frac{1}{2}(q_1 - ia_1) & \omega_3 \end{bmatrix} \quad (1)$$

The parameters of $\rho$ are related to the expectation values of expressions involving spin-1 components and their combinations.

$$\omega_i = <S_i^2> = Tr(\rho S_i^2) \quad (2a)$$
$$a_i = <S_i> = Tr(\rho S_i) \quad (2b)$$
$$q_k = <S_i S_j + S_j S_i> = Tr\{\rho(S_i S_j + S_j S_i)\}, k \neq i,j \quad (2c)$$

The invariant 3-dimensional vectors are given by the following relations [3].

(i) 1st invariant vector:
$$\vec{w} = \{\sqrt{\omega_1}, \sqrt{\omega_2}, \sqrt{\omega_3}\} \qquad (3a)$$

Based on the trace relation,
$$Tr(\rho_\Xi) = \omega_1 + \omega_2 + \omega_3 = 1, \sum_{i=1}^{3} w_i^2 = 1 \qquad (3b)$$

(ii) 2nd invariant vector:
$$\vec{u} = \{\sqrt{\omega_1^2 + (q_1^2 + a_1^2)/2}, \sqrt{\omega_2^2 + (q_2^2 + a_2^2)/2}, \sqrt{\omega_3^2 + (q_3^2 + a_3^2)/2}\} \qquad (4a)$$

Based on the trace relation
$$Tr(\rho_\Xi^2) = \sum_{i=1}^{3} u_i^2 \le 1 \qquad (4b)$$

(iii) 3rd invariant vector:
$$\vec{v} = \{\sqrt{X + 3(q_1^2 + a_1^2)/2}, \sqrt{X + 3(q_2^2 + a_2^2)/2}, \sqrt{X + 3(q_3^2 + a_3^2)/2}\} \qquad (5a)$$

Here
$$X = \frac{1}{3} - 2\omega_1\omega_2\omega_3 - \frac{1}{2}(a_2 a_3 q_1 + a_3 a_1 q_2 + a_1 a_2 q_3 - q_1 q_2 q_3) \qquad (5b)$$

Based on the trace relation
$$3Tr(\rho_\Xi^2) - 2Tr(\rho_\Xi^3) = \sum_{i=1}^{3} v_i^2 \le 1 \qquad (5c)$$

The vectors $\vec{u}$ and $\vec{v}$ represents the 2nd and 3rd density matrix invariants of a Qutrit. The bounds on the vector-norms are in general $\sum_{i=1}^{3} u_i^2 \le 1$, and $\sum_{i=1}^{3} v_i^2 \le 1$, with equality signs holding for a pure state.

For a pure state the 9 density matrix parameters are related via the following 5 relations.

$$Tr(\rho_\Xi) = \omega_1 + \omega_2 + 3 = 1 \qquad (6a)$$

$$\frac{1}{4} \begin{pmatrix} q_1^2 + a_1^2 \\ q_2^2 + a_2^2 \\ q_3^2 + a_3^2 \end{pmatrix} = \begin{pmatrix} \omega_2 \omega_3 \\ \omega_3 \omega_1 \\ \omega_1 \omega_2 \end{pmatrix} \qquad (6b)$$

$$a_2 a_3 q_1 + a_3 a_1 q_2 + a_1 a_2 q_3 - q_1 q_2 q_3 = 8\omega_1 \omega_2 \omega_3 \qquad (6c)$$

So finally we have only 4 independent parameters, which is the correct number of independent degrees of freedom for a pure Qutrit state.

III. TIME-DEPENDENT EIGENVECTORS OF THE QUTRIT CASCADE OR $\Xi$-MODEL

The isolated qutrit states with their energies are taken to be $|0> (E_0)$, $|1> (E_1)$, and $|2> (E_2)$ with $E_2 > E_1 > E_0$ average energy $\bar{E} = (E_0 + E_1 + E_2)/3$. Define, $\varepsilon_1 = (-2E_0 + E_1 + E_2)/3$, and $\varepsilon_2 = (-E_0 - E_1 + 2E_2)/3$, then the starting energies are related with them as

$$E_0 = \bar{E} - \varepsilon_1, \qquad (7a)$$

$$E_1 = \bar{E} - (\varepsilon_2 - \varepsilon_1) = \bar{E} - \varepsilon, \qquad (7b)$$

$$E_2 = \bar{E} + \varepsilon_2 \qquad (7c)$$

Here $\varepsilon = \varepsilon_2 - \varepsilon_1 = (E_0 + E_2 - 2E_1)/3 > 0$ has been assumed for later analysis ($\varepsilon < 0$ case is similar). After taking the zero of energy at $\bar{E}$, the Hamiltonian of the Qutrit in an external field is given by

$$H = \begin{bmatrix} -\varepsilon_1 & (g_1 - ig_2)\phi & (g_3 - ig_4)\phi \\ (g_1 + ig_2)\phi & -\varepsilon & (g_5 - ig_6)\phi \\ (g_3 + ig_4)\phi & (g_5 + ig_6)\phi & \varepsilon_2 \end{bmatrix} \tag{8}$$

We specialize to the $\Xi$-model in which, (i) $\varepsilon_2 = \varepsilon_1$, (ii) Couplings are equal i.e. $g_1 = g_3 = g_5, g_2 = g_4 = g_6$, and (iii) There is no coupling between $1^{st}$ and $3^{rd}$ levels. Then the Hamiltonian becomes

$$H_\Xi = \begin{bmatrix} -\varepsilon_1 & \phi(g_1 - ig_2) & 0 \\ \phi(g_1 + ig_2) & 0 & \phi(g_1 - ig_2) \\ 0 & \phi(g_1 + ig_2) & \varepsilon_1 \end{bmatrix} \tag{9}$$

Let
$$g_1 + ig_2 = Ge^{i\delta}, \tag{10a}$$
$$\omega = \sqrt{\varepsilon_1^2 + 2\phi^2 G^2} \tag{10b}$$
$$\varepsilon_1 = \omega\cos\theta \tag{10c}$$
$$\phi G\sqrt{2} = \omega\sin\theta \tag{10d}$$

Then we get

$$H_\Xi = \omega \begin{bmatrix} -\cos\theta & \frac{1}{\sqrt{2}}\sin\theta e^{-i\delta} & 0 \\ \frac{1}{\sqrt{2}}\sin\theta e^{i\delta} & 0 & \frac{1}{\sqrt{2}}\sin\theta e^{-i\delta} \\ 0 & \frac{1}{\sqrt{2}}\sin\theta e^{i\delta} & \cos\theta \end{bmatrix} \tag{11}$$

The stationary eigenvectors for eigenvalues $(-\omega, 0, \omega)$ of $H_\Xi$ are found respectively as

$$|0> = \begin{pmatrix} e^{-i\delta}\cos^2\frac{\theta}{2} \\ -\frac{1}{\sqrt{2}}\sin\theta \\ e^{i\delta}\sin^2\frac{\theta}{2} \end{pmatrix}, \quad |1> = \begin{pmatrix} \frac{1}{\sqrt{2}}e^{-i\delta}\sin\theta \\ \cos\theta \\ -\frac{1}{\sqrt{2}}e^{i\delta}\sin\theta \end{pmatrix}, \quad |2> = \begin{pmatrix} e^{-i\delta}\sin^2\frac{\theta}{2} \\ \frac{1}{\sqrt{2}}\sin\theta \\ e^{i\delta}\cos^2\frac{\theta}{2} \end{pmatrix} \tag{12}$$

The time-dependent Schrödinger's equation for the $\Xi$-model is given as

$$i\frac{\partial}{\partial t}\begin{pmatrix} |\Psi_0(t)> \\ |\Psi_1(t)> \\ |\Psi_2(t)> \end{pmatrix} = H_\Xi \begin{pmatrix} |\Psi_0(t)> \\ |\Psi_1(t)> \\ |\Psi_2(t)> \end{pmatrix} \tag{13}$$

As before, the interacting field $\phi$ is time-independent and the initial conditions are: (i) $|\psi_0(t=0)> = 1$, and (ii) $|\psi_1(t=0)> = 0 = |\psi_2(t=0)>$. The solutions for the time-dependent eigenvectors are

$$\begin{pmatrix} |\Psi_0(t)> \\ |\Psi_1(t)> \\ |\Psi_2(t)> \end{pmatrix} = \cos^2\frac{\theta}{2}e^{i(\omega t+\delta)}|0> + \frac{1}{\sqrt{2}}\sin\theta e^{i\delta}|1> + \sin^2\frac{\theta}{2}e^{-i(\omega t-\delta)}|2> \tag{14}$$

They can be rewritten as

$$\begin{pmatrix} |\Psi_0(t) > \\ |\Psi_1(t) > \\ |\Psi_2(t) > \end{pmatrix} = \begin{pmatrix} \left(1 - \frac{1}{2}\sin^2\theta\right)\cos\omega t + \frac{1}{2}\sin^2\theta + i\sin\omega t\cos\theta \\ \frac{1}{\sqrt{2}}\sin\theta[\cos\theta(1 - \cos\omega t) - i\sin\omega t]e^{i\delta} \\ -\frac{1}{2}\sin^2\theta(1 - \cos\omega t)e^{2i\delta} \end{pmatrix} \qquad (15)$$

### IV. INVARIANT VECTOR REPRESENTATION (IVR) OF A QUTRIT: Ξ-MODEL

The density matrix of the Qutrit cascade or Ξ-model is calculated as

$$\rho_\Xi = \begin{pmatrix} |\Psi_0(t) > \\ |\Psi_1(t) > \\ |\Psi_2(t) > \end{pmatrix} (< \Psi_0(t)| \quad < \Psi_1(t)| \quad < \Psi_2(t)|) \qquad (16)$$

We express the resulting density matrix as spin-1 representation given earlier. Then for the Qutrit Ξ-model, the following are the expressions for the density matrix parameters.

$$\omega_1 = \frac{1}{4}(3 + \cos 2\theta)\cos^2\omega t, \qquad (17a)$$
$$\omega_2 = \frac{1}{4}(3 + \cos 2\theta)\sin^2\omega t, \qquad (17b)$$
$$\omega_3 = \frac{1}{4}(1 - \cos 2\theta), \qquad (17c)$$

$$\begin{pmatrix} q_1 \\ a_1 \end{pmatrix} = \frac{\sin\theta}{2\sqrt{2}}(1 - \cos 2\theta)(1 - \cos\omega t)\left[\cos\theta(1 - \cos\omega t)\begin{pmatrix} \cos\delta \\ -\sin\delta \end{pmatrix} - \sin\omega t\begin{pmatrix} \sin\delta \\ \cos\delta \end{pmatrix}\right] \qquad (17d)$$

$$\begin{pmatrix} q_2 \\ a_2 \end{pmatrix} = \frac{1-\cos 2\theta}{8}(1 - \cos\omega t)\left[-\{(1 - \cos 2\theta) + (3 + \cos 2\theta)\cos\omega t\}\begin{pmatrix} \cos 2\delta \\ \sin 2\delta \end{pmatrix} + 4\cos\theta\sin\omega t\begin{pmatrix} -\sin 2\delta \\ \cos 2\delta \end{pmatrix}\right] \qquad (17e)$$

$$\begin{pmatrix} q_3 \\ a_3 \end{pmatrix} = \frac{\sin\theta}{4\sqrt{2}}\cos\theta[-(5 + 3\cos 2\theta) + 4(1 + \cos 2\theta)\cos\omega t + (1 - \cos 2\theta)\cos 2\omega t]\begin{pmatrix} \cos\delta \\ -\sin\delta \end{pmatrix} + [2(3 + \cos 2\theta)\sin\omega t + (1 - \cos 2\theta)\sin 2\omega t]\begin{pmatrix} \sin\delta \\ \cos\delta \end{pmatrix} \qquad (17f)$$

Due to the structure of the Ξ-model, there are only two independent parameters ($\omega, \theta$). Using the earlier expressions, the IVR vectors for Ξ-model are found to be

(i) 1$^{st}$-order invariant vector ($\vec{w} = \{\sqrt{\omega_1}, \sqrt{\omega_2}, \sqrt{\omega_3}\}$): The angles in spherical representation are:

Colatitude angle or the angle between the IVR vector and z-axis
$$\psi_1 = \cos^{-1}(\sqrt{\omega_3}) = \cos^{-1}\left(\frac{1}{\sqrt{2}}\sin\theta\right), \qquad (18a)$$
It is time-independent,

Azimuthal angle or the angle between projection of IVR on XY-plane and the x-axis
$$\chi_1 = \tan^{-1}\sqrt{\frac{\omega_2}{\omega_1}} = \omega t \text{ (modulo } 2\pi\text{)}, \qquad (18b)$$
It has linear time-dependence.

(ii) 2$^{nd}$-order invariant vector ($\vec{u} = \{\sqrt{\omega_1^2 + 2\omega_2\omega_3}, \sqrt{\omega_2^2 + 2\omega_3\omega_1}, \sqrt{\omega_3^2 + 2\omega_1\omega_2}\}$):

The IVR angles for $\vec{u}(\psi_2, \chi_2)$ are:

Colatitude angle or the angle between the IVR vector and Z-axis:
$$\psi_2 = \cos^{-1}\sqrt{\omega_3^2 + 2\omega_1\omega_2}, \qquad (19a)$$

Azimuthal angle or the angle between projection of IVR on XY-plane and the X-axis

$$\chi_2 = tan^{-1}\sqrt{\frac{\omega_2^2+2\omega_3\omega_1}{\omega_1^2+2\omega_2\omega_3}} \qquad (19b)$$

We calculate and plot $(\psi_2, \omega t)$ and $(\chi_2, \omega t)$ for values of $\theta = 3.0$ and $5.0$ as shown in Fig 1-4.

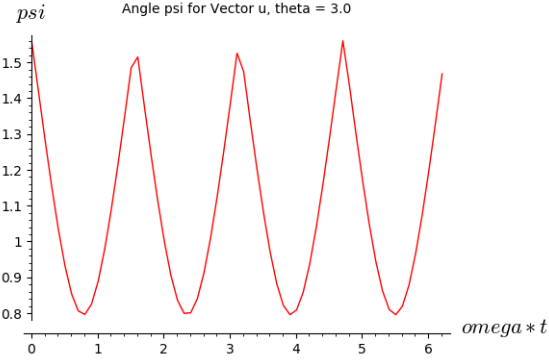

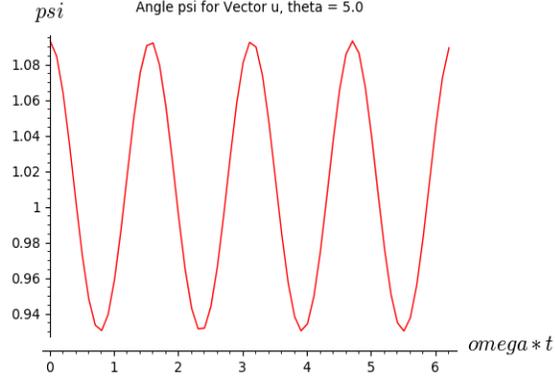

Fig.1: Colatitude angle of the 2$^{nd}$ order Invariant Vector u as a function of $\omega t$ for $\theta$=3 radians

Fig.2: Colatitude angle of the 2$^{nd}$ order Invariant Vector u as a function of $\omega t$ for $\theta$=5 radians

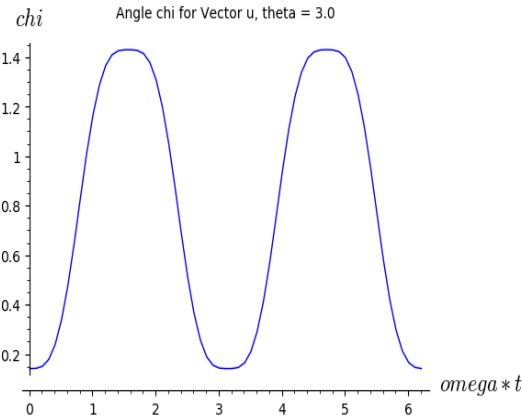

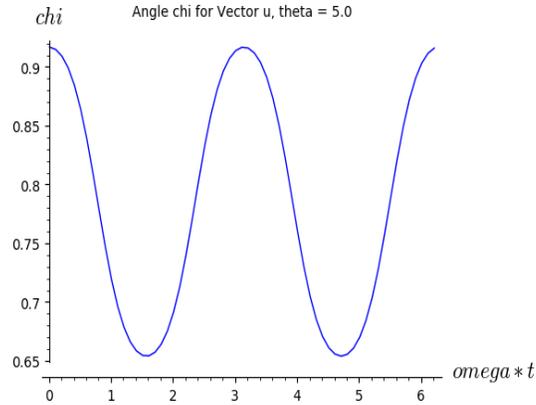

Fig.3: Azimuthal angle of the 2$^{nd}$ order Invariant Vector u as a function of $\omega t$ for $\theta$=3 radians

Fig.4: Azimuthal angle of the 2$^{nd}$ order Invariant Vector u as a function of $\omega t$ for $\theta$=5 radians

These angles show sinusoidal-like variations of the spherical angles of IVR vectors for two widely varying values of undressed energy given by $\varepsilon_1 = \omega cos\theta$. It is conjectured that similar behavior will be also present in the $\vec{u}(\psi_2, \chi_2)$ of Qutrit $\Lambda$ and V models.

(iii) 3$^{rd}$-order invariant vector: $\vec{v} = \left\{\sqrt{\frac{1}{3}}, \sqrt{\frac{1}{3}}, \sqrt{\frac{1}{3}}\right\}$

The 3$^{rd}$ order vector turns out to be a constant one for the $\Xi$-model and angles are

$$\psi_3 = \cos^{-1}\sqrt{\frac{1}{3}} \cong 55^0 \tag{20a}$$

$$\chi_3 = \tan^{-1}(1) = 45^0 \tag{20b}$$

The spherical angles associated with other 2 vectors $\vec{w}(\psi_1, \chi_1)$ and $\vec{u}(\psi_2, \chi_2)$ capture the time-dependent dynamic essence of this model.

## V. MSR AND IVR DENSITY MATRICES OF A QUTRIT

An alternative method for representing qutrit is the Majorana Star Representation (MSR) based on multi-qubit spin states [7, 8, 10, 11]. In MSR a qutrit is represented by two unit length vectors on a Bloch sphere. The IVR vectors also reside on the surface of a sphere but that is not a Bloch sphere. In this section both representations for pure states are compared and the general transformation between them is also presented.

We choose two points $P_1(\theta_1, \phi_1)$ and $P_2(\theta_2, \phi_2)$ or equivalently $P_1(x_1, y_1, z_1)$ and $P_2(x_2, y_2, z_2)$ on the Bloch sphere centered at point $O$. Then the Cartesian and spherical components of these Majorana stars for qutrit are given as

$$\overrightarrow{OP_1} = (x_1, y_1, z_1) = (\sin\theta_1\cos\phi_1, \sin\theta_1\sin\phi_1, \cos\theta_1) \tag{21a}$$

$$\overrightarrow{OP_2} = (x_2, y_2, z_2) = (\sin\theta_2\cos\phi_2, \sin\theta_2\sin\phi_2, \cos\theta_2) \tag{21b}$$

Let the general qutrit state be given as

$$\begin{pmatrix} C_1 \\ C_0 \\ C_{-1} \end{pmatrix} = C_1\begin{pmatrix}1\\0\\0\end{pmatrix} + C_0\begin{pmatrix}0\\1\\0\end{pmatrix} + C_{-1}\begin{pmatrix}0\\0\\1\end{pmatrix} \tag{22}$$

Then the qutrit state coefficients are related by Majorana's Star equation specialized to a qutrit.

$$\frac{C_1}{\sqrt{2}}\varsigma^2 - C_0\varsigma + \frac{C_{-1}}{\sqrt{2}} = 0 \tag{23}$$

The solutions of this quadratic equation are

$$\begin{pmatrix} \varsigma_1 \\ \varsigma_2 \end{pmatrix} = \frac{1}{C_1\sqrt{2}}\left[C_0\begin{pmatrix}+\\-\end{pmatrix}\sqrt{C_0^2 - 2C_1C_{-1}}\right] \tag{24}$$

These solutions are related to the starting Majorana star vectors through the following steps.

*Step 1*: The inverse stereographic projections of $P_1(\theta_1, \phi_1)$ and $P_2(\theta_2, \phi_2)$ give the two points lie on the equatorial plane of the Bloch sphere. Their Cartesian components are related to those of the stars.

$$\begin{pmatrix}\alpha_1\\\alpha_2\end{pmatrix} = \begin{pmatrix}x_1/(1+z_1)\\x_2/(1+z_2)\end{pmatrix}, \tag{25a}$$

$$\begin{pmatrix}\beta_1\\\beta_2\end{pmatrix} = \begin{pmatrix}y_1/(1+z_1)\\y_2/(1+z_2)\end{pmatrix}, \tag{25b}$$

*Step 2*: The coordinates are combined to give complex quantities, which are solutions to the Majorana Star equation.

$$\begin{pmatrix} \varsigma_1 \\ \varsigma_2 \end{pmatrix} = \begin{pmatrix} \alpha_1 + i\beta_1 \\ \alpha_2 + i\beta_2 \end{pmatrix} = \begin{pmatrix} \exp(i\phi_1)\tan\left(\frac{\theta_1}{2}\right) \\ \exp(i\phi_2)\tan\left(\frac{\theta_2}{2}\right) \end{pmatrix} \qquad (26)$$

Then the pure qutrit state coefficients are given by

$$\begin{pmatrix} C_1 \\ C_0 \\ C_{-1} \end{pmatrix} = \frac{1}{N} \begin{pmatrix} 1 \\ (\varsigma_1 + \varsigma_2)/\sqrt{2} \\ \varsigma_1 \varsigma_2 \end{pmatrix}, \qquad (27a)$$

The normalization is

$$N = \left( \frac{(1+\cos\theta_1)(1+\cos\theta_2)}{3 + 4\cos\left(\frac{\theta_1}{2}\right)\cos\left(\frac{\theta_2}{2}\right) + \sin\left(\frac{\theta_1}{2}\right)\sin\left(\frac{\theta_2}{2}\right)\cos(\phi_1-\phi_2)} \right)^{1/2} \qquad (27b)$$

The MSR qutrit density matrix then becomes

$$\rho_{MSR} = \begin{pmatrix} C_1 \\ C_0 \\ C_{-1} \end{pmatrix} \begin{pmatrix} \overline{C_1} & \overline{C_0} & \overline{C_{-1}} \end{pmatrix} = \frac{1}{N^2} \begin{pmatrix} 1 & (\overline{\varsigma_1} + \overline{\varsigma_2})/\sqrt{2} & \overline{\varsigma_1}\overline{\varsigma_2} \\ (\varsigma_1 + \varsigma_2)/\sqrt{2} & |\varsigma_1 + \varsigma_2|^2/2 & \overline{\varsigma_1}\overline{\varsigma_2}(\varsigma_1 + \varsigma_2)/\sqrt{2} \\ \varsigma_1\varsigma_2 & \varsigma_1\varsigma_2(\overline{\varsigma_1} + \overline{\varsigma_2})/\sqrt{2} & |\varsigma_1\varsigma_2|^2 \end{pmatrix} \qquad (28)$$

The density matrix $\rho$ based on the IVR of a Qutrit has been given earlier in eqn.(1). The parameters of the density matrix $\rho_{IVR}$ are related to the expectation values of expressions involving spin-1 components and their combinations as given in eqn.(2). The qutrit in IVR is represented by 3 vectors as given in the following table.

Table-1: The IVR vectors for qutrit

| IVR vectors | Cartesian components | Basis trace relations | Comments |
| --- | --- | --- | --- |
| $\vec{w}$ | $\sqrt{\omega_1}, \sqrt{\omega_2}, \sqrt{\omega_3}$ | $\sum_{i=1}^{3} w_i^2 = Tr(\rho_\Xi) = \omega_1 + \omega_2 + \omega_3 = 1$ | Follows from unit trace relation of density matrix |
| $\vec{u}$ | $\sqrt{\omega_1^2 + (q_1^2 + a_1^2)/2}$, $\sqrt{\omega_2^2 + (q_2^2 + a_2^2)/2}$, $\sqrt{\omega_3^2 + (q_3^2 + a_3^2)/2}$ | $Tr(\rho_\Xi^2) = \sum_{i=1}^{3} u_i^2 \leq 1$ | Trace idempotence relation for pure states |
| $\vec{v}$ | $\sqrt{X + 3(q_1^2 + a_1^2)/2}$, $\sqrt{X + 3(q_2^2 + a_2^2)/2}$, $\sqrt{X + 3(q_3^2 + a_3^2)/2}$ | $3Tr(\rho_\Xi^2) - 2Tr(\rho_\Xi^3) = \sum_{i=1}^{3} v_i^2 \leq 1$ | Here $X = \frac{1}{3} - 2\omega_1\omega_2\omega_3$ $-\frac{1}{2}(a_2 a_3 q_1 + a_3 a_1 q_2 + a_1 a_2 q_3 - q_1 q_2 q_3)$ |

The vectors $\vec{u}$ and $\vec{v}$ represents the 2nd and 3rd density matrix invariants of a Qutrit. The bounds on the vector-norms are in general $\sum_{i=1}^{3} u_i^2 \leq 1$, and $\sum_{i=1}^{3} v_i^2 \leq 1$, with equality signs holding for a pure state. For a pure state the 8 density matrix parameters are related via the following 4 relations, (i) $q_1^2 + a_1^2 = 4\omega_2\omega_3$, (ii) $q_2^2 + a_2^2 = 4\omega_3\omega_1$, (iii) $q_3^2 + a_3^2 = 4\omega_1\omega_2$, and (iv) $a_2 a_3 q_1 + a_3 a_1 q_2 + a_1 a_2 q_3 - q_1 q_2 q_3 = 8\omega_1\omega_2\omega_3$. So finally we have only 4 independent parameters, which is the correct number of independent degrees of freedom for a pure Qutrit state.

## VI. RELATION BETWEEN THE MSR AND IVR OF A QUTRIT

We equate the two density matrices representing the same qutrit entity.

$$\rho_{IVR} = \rho_{MSR} \tag{29}$$

Then the density matrix elements of the MSR and IVR are related by,

$$\omega_1 = \frac{1}{N^2}, \tag{30a}$$

$$\omega_2 = \frac{1}{2N^2}[(\alpha_1 + \alpha_2)^2 + (\beta_1 + \beta_2)^2], \tag{30b}$$

$$\omega_3 = \frac{1}{N^2}(\alpha_1^2 + \beta_1^2)(\alpha_2^2 + \beta_2^2) \tag{30c}$$

$$\begin{pmatrix} q_1 \\ a_1 \end{pmatrix} = -\frac{\sqrt{2}}{N^2} \begin{bmatrix} \alpha_1(\alpha_2^2 + \beta_2^2) + \alpha_2(\alpha_1^2 + \beta_1^2) \\ \beta_1(\alpha_2^2 + \beta_2^2) + \beta_2(\alpha_1^2 + \beta_1^2) \end{bmatrix}, \tag{31}$$

$$\begin{pmatrix} q_2 \\ a_2 \end{pmatrix} = \frac{2}{N^2} \begin{pmatrix} \alpha_1\alpha_2 - \beta_1\beta_2 \\ \alpha_1\beta_2 + \alpha_2\beta_1 \end{pmatrix}, \tag{32}$$

$$\begin{pmatrix} q_3 \\ -a_3 \end{pmatrix} = \frac{\sqrt{2}}{N^2} \begin{pmatrix} \alpha_1 + \alpha_2 \\ \beta_1 + \beta_2 \end{pmatrix} \tag{33}$$

The invariant vectors have the following expressions.

$$(i) \vec{w} = \{\sqrt{\omega_1}, \sqrt{\omega_2}, \sqrt{\omega_3}\} = (sin\psi_1 cos\chi_1, sin\psi_1 sin\chi_1, cos\psi_1) \tag{34}$$

We use the pure qutrit relations given earlier to get the following forms for the other two vectors.

$$(ii) \vec{u} = \{\sqrt{\omega_1^2 + 2\omega_2\omega_3}, \sqrt{\omega_2^2 + 2\omega_3\omega_1}, \sqrt{\omega_3^2 + 2\omega_1\omega_2}\}$$
$$= (sin\psi_2 cos\chi_2, sin\psi_2 sin\chi_2, cos\psi_2) \tag{35}$$

$$(iii) \vec{v} = \left\{\sqrt{\frac{1}{3}}, \sqrt{\frac{1}{3}}, \sqrt{\frac{1}{3}}\right\} = (sin\psi_3 cos\chi_3, sin\psi_3 sin\chi_3, cos\psi_3) \tag{36}$$

The vector $\vec{v}$ reduces to a constant vector for a pure state. This form of 3rd order invariant vector is conjectured to be an indicator of a pure qutrit state.

## VII. MSR AND IVR VECTORS OF THE QUTRIT CASCADE MODEL

One can study the relationship between $(\theta_1, \phi_1, \theta_2, \phi_2)$ and $(\psi_1, \chi_1, \psi_i, \chi_2)$ sets in many ways to map the MSR vectors into IVR ones and vice versa. The plots given below show the relationship between the angles for a particular case in which IVR angles are calculated for the following MSR case:
  (i) The MSR $\overrightarrow{OP_1}$ vector is fixed ($\theta_1$=1 rad, $\phi_1$=1 rad).
  (ii) The azimuth angle of the MSR $\overrightarrow{OP_2}$ vector is fixed ($\phi_2$=4 rad).

The plots show the behavior of the co-latitude angles ($\psi_1, \psi_2$) and azimuth angles ($\chi_1, \chi_2$) of the IVR vectors $\vec{w}$ and $\vec{u}$ respectively.

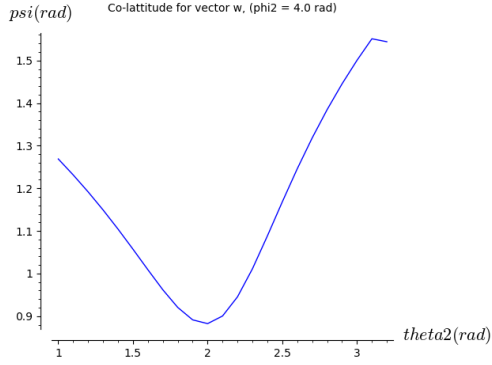

Fig.1: Colatitude angle of the 1st order Invariant Vector $\vec{w}$ as a function of MSR colatitude angle $\theta_2$ of $\overrightarrow{OP_2}$ ($\phi_2$ is fixed).

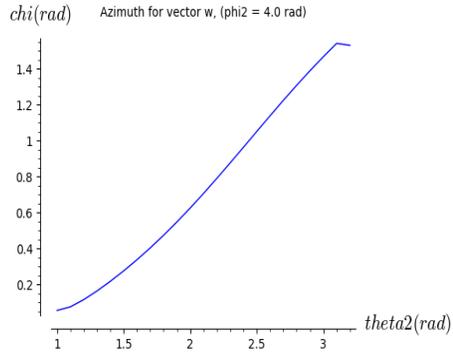

Fig.2: Azimuth angle of the 1st order Invariant Vector $\vec{w}$ as a function of MSR colatitude angle $\theta_2$ of $\overrightarrow{OP_2}$ ($\phi_2$ is fixed).

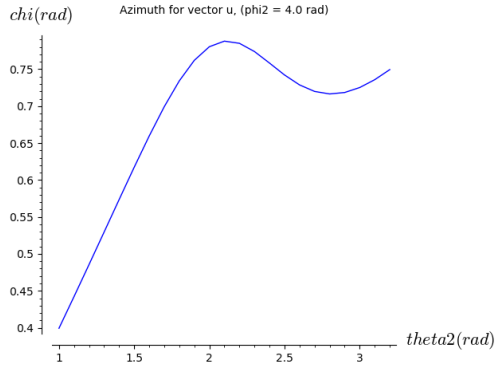

Fig.3: Colatitude angle of the 2nd order Invariant Vector $\vec{u}$ as a function of MSR colatitude angle $\theta_2$ of $\overrightarrow{OP_2}$ ($\phi_2$ is fixed).

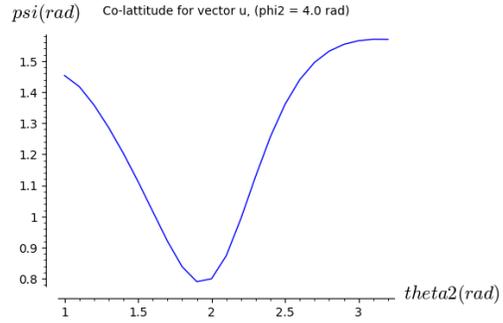

Fig.4: Azimuth angle of the 2nd order Invariant Vector $\vec{u}$ as a function of MSR colatitude angle $\theta_2$ of $\overrightarrow{OP_2}$ ($\phi_2$ is fixed).

It is seen that the two vector sets are related nonlinearly, which was to be expected given the complicated relations between them in the two representations. In practice it is easier to find the IVR from given MSR and inverse relations are much harder to calculate.

## VIII. SUMMARY AND NEXT STEPS

The 3-dimensional IVR vectors representing qutrit states have been shown to capture the essential dynamics of cascade or Ξ-model. The model qutrit state is pure and so has fewer degrees of freedom. Out of three vectors only one was found to display complex behavior. Out of the remaining two vectors, one is a constant and another has a linearly time dependent azimuthal angle..

On the other hand, the IVR is capable of displaying the full static or dynamic behavior of a mixed qutrit state with all 8 degrees of freedom as well. In that situation the qutrit dynamics is expressed by the behavior of 3 parameters each of $\vec{u}$ and $\vec{v}$ and 2 parameters of $\vec{w}$ as it has unit length by definition. This is a significant advance compared with traditional approach based on Gell-Mann SU(3) matrices. In future, the IVR will be applied to study the relative performance of different qutrit models for Quantum Sensing. One can also see that given the angular variables of Majorana Stars for a qutrit, it is very straightforward to calculate the corresponding IVR angles and vice versa. They are related in a complicated manner as seen by the plots given earlier.

The IVR representation is more versatile as it can also represent mixed qutrit states, which cannot be represented by Majorana Stars. Recently MSR has been extended to mixed states [11] so IVR may be connected to this extended MSR.